\documentclass[prd,twocolumn,a4paper,superscriptaddress,floatfix]{revtex4}
\usepackage{graphicx}
\usepackage{bbm}
\usepackage[all]{xy}
\usepackage{amsmath}
\usepackage{amssymb}
\usepackage{epstopdf}

\def\be{\begin{equation}}
\def\ee{\end{equation}}
\newcommand{\bq}{\begin{eqnarray}}
\newcommand{\eq}{\end{eqnarray}}
\newcommand{\bes}{\begin{subequations}}
\newcommand{\ees}{\end{subequations}}

\def\ben{\begin{eqnarray}}
\def\een{\end{eqnarray}}
\def\ba{\begin{array}}
\def\ea{\end{array}}

\begin{document}
\newcommand{\half}{{\textstyle\frac{1}{2}}}
\allowdisplaybreaks[3]
\def\a{\alpha}
\def\b{\beta}
\def\g{\gamma}\def\G{\Gamma}
\def\d{\delta}\def\D{\Delta}
\def\ep{\epsilon}
\def\et{\eta}
\def\z{\zeta}
\def\t{\theta}\def\T{\Theta}
\def\l{\lambda}\def\L{\Lambda}
\def\m{\mu}
\def\f{\phi}\def\F{\Phi}
\def\n{\nu}
\def\p{\psi}\def\P{\Psi}
\def\r{\rho}
\def\s{\sigma}\def\S{\Sigma}
\def\ta{\tau}
\def\x{\chi}
\def\o{\omega}\def\O{\Omega}
\def\k{\kappa}
\def\pa {\partial}
\def\ov{\over}
\def\br{\\}
\def\ud{\underline}

\newcommand\lsim{\mathrel{\rlap{\lower4pt\hbox{\hskip1pt$\sim$}}
    \raise1pt\hbox{$<$}}}
\newcommand\gsim{\mathrel{\rlap{\lower4pt\hbox{\hskip1pt$\sim$}}
    \raise1pt\hbox{$>$}}}
\newcommand\esim{\mathrel{\rlap{\raise2pt\hbox{\hskip0pt$\sim$}}
    \lower1pt\hbox{$-$}}}

\title{Domain wall network evolution in $(N+1)$-dimensional FRW universes}

\author{P.P. Avelino}
\email[Electronic address: ]{ppavelin@fc.up.pt}
\affiliation{Centro de F\'{\i}sica do Porto, Rua do Campo Alegre 687, 4169-007 Porto, Portugal}
\affiliation{Departamento de F\'{\i}sica da Faculdade de Ci\^encias
da Universidade do Porto, Rua do Campo Alegre 687, 4169-007 Porto, Portugal}
\affiliation{Departamento de F\'{\i}sica, Universidade Federal da Para\'{\i}ba
58051-970 Jo\~ao Pessoa, Para\'{\i}ba, Brazil}
\author{L. Sousa}
\email[Electronic address: ]{laragsousa@gmail.com}
\affiliation{Centro de F\'{\i}sica do Porto, Rua do Campo Alegre 687, 4169-007 Porto, Portugal}
\affiliation{Departamento de F\'{\i}sica da Faculdade de Ci\^encias
da Universidade do Porto, Rua do Campo Alegre 687, 4169-007 Porto, Portugal}

\begin{abstract}

We develop a velocity-dependent one-scale model for the evolution of domain wall networks in flat expanding or collapsing homogeneous and isotropic universes with an arbitrary number of spatial dimensions, finding the corresponding scaling laws in frictionless and friction dominated regimes. We also determine the allowed range of values of the curvature parameter and the expansion exponent for which a linear scaling solution is possible in the frictionless regime.

\end{abstract} 
\pacs{98.80.Cq}
\maketitle

\section{Introduction}

Domain walls can form at a spontaneous phase transition with discrete symmetry breaking in the early universe \cite{1994csot.book.....V}. Their evolution after the phase transition  \cite{Press,Sousa:2010zz} plays an important part in determining their possible cosmic role. Although, current observational constraints on the dark energy equation of state parameter strongly disfavor domain walls as the single dark energy component \cite{Frieman:2008sn,Komatsu:2010fb}, such limits, on their own, do not yet rule out a substantial impact of a frustrated domain wall network  \cite{Bucher:1998mh} on the acceleration of the Universe around the present time.

The dynamics and cosmological consequences of domain wall networks in flat, expanding or collapsing homogeneous and isotropic universes with $3+1$ dimensions have been extensively studied in the literature, using both semi-analytical models \cite{Avelino:2005kn,PinaAvelino:2006ia,Sousa:2009is}  and domain wall network simulations \cite{Avelino:2006xy,Battye:2006pf,Avelino:2006xf,Avelino:2008ve,Avelino:2009tk} considering complex models with Y- and/or X-type junctions \cite{Kubotani:1991kw,Carter:2004dk,Bazeia:2005wt}. These studies, in combination with recent observational results, have provided very strong evidence that domain walls cannot account for a significant fraction of the dark energy.

Until the present date, analytical studies of domain wall dynamics in more than $3$ spatial dimensions have only considered the evolution of maximally symmetric configurations \cite{Avelino:2008mv}. Although a number of interesting results have been obtained, they cannot be directly applied to the evolution of domain wall networks in homogeneous and isotropic universes with an arbitrary number of spatial dimensions. In this paper we eliminate this shortcoming by generalizing the existing velocity-dependent one-scale (VOS) model, describing the evolution of the characteristic length and velocity of a domain wall network, to an arbitrary number of spatial dimensions.

The paper is organized as follows. In Sec. II we develop a VOS model for the evolution of $(N-1)$-branes in $(N+1)$-dimensional Friedmann-Robertson-Walker (FRW) universes. In Sec. III we study the frictionless regime, finding a number of scaling solutions describing the dynamics of domain walls in in $(N+1)$-dimensional expanding or collapsing FRW universes. In section IV we investigate the friction dominated regimes. We then conclude in Sec. V. Throughout the work, we will assume the metric signature $[+,-,...,-]$ and the calculations will be done using units in which $c=\hbar=1$. 

\section{Velocity-dependent One Scale Model for $(N-1)$-branes}

In a $(N+1)$-dimensional FRW Universe, the dynamics of a featureless $(N-1)$-brane whose thickness is much smaller than its curvature radii is described by \cite{Sousa:2010zz}
\be
{\dot v}+\left(1-v^2\right)\left[NHv-\kappa \right]=0\,,
\label{dyn1}
\ee
where  a dot represents a derivative with respect to physical time $t$, $a$ is the scale factor, $H={\dot a}/a$ is the Hubble parameter, $v$ is the microscopic velocity of the surface and $\kappa$ is the curvature of the surface defined at each point as the sum of the principal curvatures.

Now consider the case of a network of $(N-1)$-branes (domain walls) in a $(N+1)$-dimensional FRW Universe and let us start by defining the r.m.s velocity ${\bar v}={\sqrt {\langle v^2\rangle}}$ as
\be
{\bar v}^2=\frac{\int v^2   \rho \,  dV}{\int \rho \,  dV}\,,
\ee
where $\rho$ is the domain wall energy density and $V$ is the physical volume. Let us also define the characteristic length, $L$, of the domain wall network as 
\be
{\bar \rho}=\frac{\sigma_{N-1}}{L}\,,
\ee 
where $\sigma_{N-1}$ is the wall energy per unit $N-1$-dimensional area and ${\bar \rho}=V^{-1}\int \rho \, dV$ is the average domain wall density.

Multiplying Eq. (\ref{dyn1}) by $v$, making the above weighted volume average and then dividing by ${\bar v}$ one obtains
\be
{\dot {\bar v}}+\frac{1}{\bar v} \langle v \left(1-v^2\right)\left[NHv-\kappa \right] \rangle=0\,.
\label{dyn2}
\ee
If we assume that $\langle v^4 \rangle={\bar v}^4$ then Eq. (\ref{dyn2}) further simplifies to
\be
{\dot {\bar v}}+\left(1-{\bar v}^2\right)\left[\frac{\bar v}{\ell_{\rm d}}-\frac{k}{L}\right]=0\,,
\label{dyn3}
\ee
where $\ell_{\rm d}^{-1}=NH$, $k={\bar \kappa} L$ and
\be
{\bar \kappa}=\frac{\langle v(1-v^2) \kappa \rangle}{{\bar v} (1-{\bar v}^2)} = \frac{\int v \, (1-v^2) \,  \kappa \,  \rho \, dV}{{\bar v}(1-{\bar v}^2)\int \rho \, dV}\,.
\ee
The above assumption is valid in the relativistic regime up to first order in $(1-v)$ and it has a negligible impact in the non-relativistic limit. Although it is possible to construct network configurations with the same $\bar v$ but different ${\bar \kappa}$, in most physically realistic situations it is sufficient to consider that ${\bar \kappa}={\bar \kappa}({\bar v})$ \cite{Martins:2000cs}. 
In the presence of frictional forces the characteristic damping length becomes $\ell_{\rm d}^{-1}=NH+\ell_{\rm f}^{-1}$, where $\ell_{\rm f}$ is the friction lengthscale.

Let us assume that the domain wall network is statiscally homogeneous and isotropic on large enough scales, so that it behaves as a brane gas. Energy-momentum conservation in a FRW universe implies that
\be
{\dot {\bar \rho}}+NH\left({\bar \rho}+{\bar {\mathcal P}}\right)=0\,.
\label{em-frw}
\ee
where ${\bar {\mathcal P}}=V^{-1}\int {\mathcal P} \, dV $ and ${\bar \rho}=V^{-1}\int \rho \, dV$ are the average domain wall pressure and density whose ratio is given by the equation of state parameter \cite{Boehm:2002bm}
\be
w=\frac{{\bar {\mathcal P}}}{\bar \rho}=\left({\bar v}^2-\frac{N-1}{N}\right)\,.
\label{eos}
\ee
Eq. (\ref{em-frw}) can be generalized in order to account for additional energy loss mechanisms, due to interface collapse and friction
\be
{\dot {\bar \rho}}+NH\left({\bar \rho}+{\bar {\mathcal P}}\right)=-\left(\frac {{\tilde c}{\bar v}}{L}+\frac{{\bar v}^2}{\ell_{\rm f}}\right){\bar \rho}\,.
\label{enc1}
\ee
Here, ${\tilde c}$ is a phenomenological parameter which depends on the specific properties of the network and can be calibrated 
using numerical simulations  \cite{Avelino:2008ve}. Using Eq.  (\ref{enc1}) one obtains 
\be
{\dot L}=HL+\frac{L}{\ell_{\rm d}} {\bar v}^2+{\tilde c}{\bar v}\,.
\label{vos2}
\ee
Eqs. (\ref{dyn3}) and (\ref{vos2}) constitute a generalization of the VOS model in \cite{Avelino:2005kn} to the case of a $N+1$-dimensional FRW Universe, which has been derived here in detail.

\section{Frictionless regime}

For simplicity, we shall assume that the dynamics of the universe is driven by a fluid with $w={\rm constant} \neq -1$ so that $a \propto t_*^\beta$, where $\beta=2/(N(w+1))$ and $t_* \ge 0$ is the time elapsed since the initial singularity (if $dt_*=dt$) or the time remaining up to the final singularity (if $dt_*=-dt$) at $t_*=0$. We shall label the various models as $M^s_i$. Here, $s=\pm$ depending on whether $dt=\pm dt_*$ and $i=1,2$ or $3$,  depending on whether $\beta <0$, $0<\beta <1$ or 
$\beta >1$, respectively. We shall also assume that ${\tilde c} \ge 0$. The models $M^+_2$, $M^+_3$ and $M^-_1$ represent expanding solutions with $t_*=0$ either at the the big-bang ($M^+_2$ and $M^+_3$) or at the big rip (for $M^-_1$). The models $M^+_1$, $M^-_2$ and $M^-_3$ represent collapsing  universes with $t_*=0$ either at the the big-crunch ($M^-_2$ and $M^-_3$) or at an initial infinite density singularity with $a(t_*=0)=\infty$ (for $M^+_1$). In an expanding universe, if $w > -1$, the Hubble radius, $|H|^{-1}$, increases with time but for $-1< w < w_c$ (with $w_c=(2-N)/N$, so that $\beta > 1$) the comoving Hubble radius ($|H|^{-1}/a$) decreases with time. Note that, in an expanding universe, the comoving Hubble radius will monotously increase or decrease with time depending on whether $\ddot{a}$ is negative or positive, respectively. Of course, the reverse is true in a collapsing regime.

\renewcommand{\arraystretch}{1.4}
\begin{table}[h]
\begin{center}
\begin{tabular}{|c|c|c|c|c|c|} \hline
 &  $\dot{a}$  &  $\ddot{a}$  &  $\dot{H}$  & C/E & Main Characteristic\\ \hline\hline
$M_1^+$ & $-$ & $+$ & $+$ & C & Linear Scaling  \\ \hline
$M_2^+$ & $+$ & $-$ & $-$ & E & Linear Scaling \\ \hline
$M_3^+$ & $+$ & $+$ & $-$ & E & Inflation \\ \hline
$M_1^-$ & $+$ & $+$ & $+$ & E & SuperInflation \\ \hline
$M_2^-$ & $-$ & $-$ & $-$ & C & Ultrarelativistic \\ \hline
$M_3^-$ & $-$ & $+$ & $-$ & C & Linear Scaling \\ \hline
\end{tabular}
\end{center}
\caption{Summary of the main properties of the different $M_i^s$ models. The label $i$ takes the value $i=1,2 \,\, \mbox{or}\,\,3$ for  $\beta<0$, $0<\beta<1$ or $\beta>1$, respectively. On the other hand, $s=\pm$ depending on whether $dt=\pm dt_*$ and $t_*$ is the time elapsed since the initial singularity ($s=+$) or the time remaining until the final singularity ($s=-$). The remaining  $+$ and $-$ indicate the sign of the cosmological parameters represented in the table and the letters $C$ and $E$ represent collapse and expansion, respectively.}
\end{table}

\begin{figure}[htb]
\includegraphics[width=3.5in]{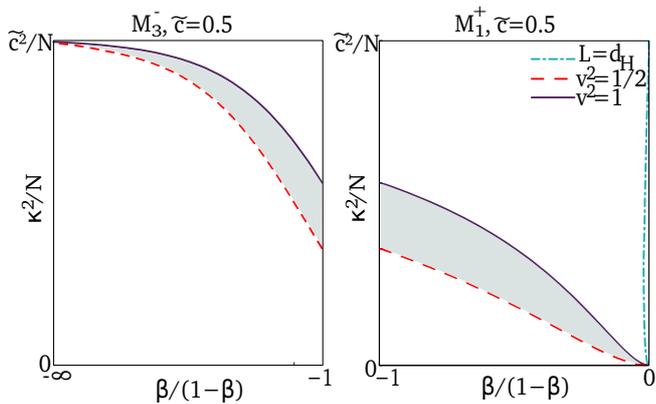}
\caption{\label{cc} The range of values of the curvature parameter, $k$, for which the VOS equations admit a linear scaling solution (grey area), as a function of  $\beta/(1-\beta)$, for the models $M_3^-$ (left panel) and $M_1^+$ (right panel), with ${\tilde c}=0.5$. The dash-dotted (blue), dashed (red) and solid (purple) lines are defined by $L=d_H$, ${\bar v}^2=1/2$ and ${\bar v}^2=1$, respectively.}
\end{figure}

\subsection{Linear scaling solutions ($\ell_{\rm f}=\infty$)}

If the friction lengthscale becomes negligible compared to the Hubble radius then Eqs. (\ref{dyn3}) and (\ref{vos2}) may have a linear attractor solution. This attractor solution corresponds to a linear scaling regime of the form
\be
L=\xi t_* \,,  \qquad \bar v=\mbox{const}\,.
\ee
so that
\be
{\bar v}^2=\frac{(1-\beta)k}{N\beta(k+{\tilde c})}\,, \qquad \xi=\pm \frac{k}{N\beta v}\,,
\ee
and, finally, 
\be
\xi= {\sqrt {\left| {\frac{k(k+{\tilde c})}{N \beta (1-\beta)}}\right|}}\,, \qquad {\bar v}={\sqrt { {\frac{1-\beta}{N\beta}\frac{k}{k+{\tilde c}}}}}\,. \label{scaling}
\ee
Two necessary conditions for a linear scaling solution of the VOS equations to be possible are
\be
0 < {\bar v} <1\,, \qquad \xi > 0\,,
\ee
but these are by no means sufficient. There are a number of complementary constraints which significantly reduce the range of parameters consistent with a linear scaling solution.

The r.m.s. velocity of maximally symmetric $(N-1)$-branes with a $S_{N-1-i}\otimes \mathbbm{R}^i$ topology oscillating periodically in a Minkowski spacetime is given by \cite{Avelino:2008mv}
\be
{\bar v}^2 = \frac{N-1-i}{N-i}\label{averagevel}\,.
\ee
The value of ${\bar v}^2$ takes the maximum value, equal to $1-1/N$, if $i=0$ (i.e., all the principal curvatures of the surface are equal and non-zero) and the minimum value, equal to $1/2$, if $i=N-2$ (i.e., if only one of the principal curvatures is non-zero). Causality constraints do not allow for infinite flat branes to be formed in realistic cosmological scenarios and consequently we did not consider the case with $i=N-1$, where all the principal curvatures are equal to zero, in the above discussion.

\begin{figure}[htb]
\includegraphics[width=3.5in]{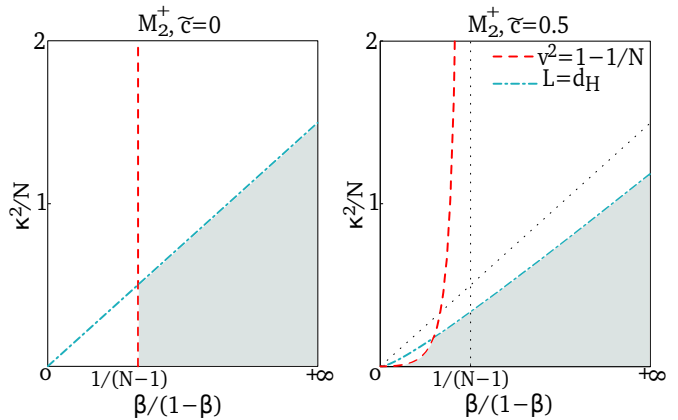}
\caption{\label{ce} The range of values of the curvature parameter, $k$, for which the VOS equations admit a linear scaling solution, as a function of  $\beta/(1-\beta)$, for the $M^+_2$ model (grey area). The left panel represents the allowed range of $k^2/N$ with ${\tilde c}=0$ and the right panel shows how this range would be changed for ${\tilde c}=0.5$. 
The dash-dotted (blue) and dashed (red) are defined by $L=d_H$ and $v^2={\bar v}^2_{\rm max}=1-1/N$, respectively}
\end{figure}

We may use this result to estimate the minimum and maximum r.m.s. velocity of a domain wall network in a $(N+1)$-dimensional Minkowski spacetime (corresponding to $\beta=0$) as 
\be
v^2_{\rm min}=\frac{1}{2}\,, \qquad v^2_{\rm max}=1-\frac{1}{N}\,.
\ee
Note that for $\beta=0$ the curvature parameter $k$ must be equal to zero for a linear scaling solution with ${\bar v} \le 1$ to be attained. The expansion (collapse) of the universe will add a damping (forcing)  which will necessarily lead to a smaller (larger) r.m.s. velocity, ${\bar v}$ and a curvature parameter, $k$, larger (smaller) than zero.

Also, causality constraints require the characteristic length of the network to be smaller than the particle horizon, $d_H$, at any given time
\be
L < d_H = \int_{t_i}^t \frac{dt'}{a(t')}\label{causality}\,,
\ee
where $t_i=0$ or $t_i=-\infty$ depending on whether $s=+$ or $-$, respectively. The particle horizon is infinite in the case of the 
models $M_3^+$, $M_1^-$ and $M_2^-$. The constraint given by Eq. (\ref{causality}) is only relevant in the case of the 
models for which a linear scaling solution is possible ($M_1^+$, $M_2^+$ and $M_3^-$) and can be written as $\xi<\left|1-\beta\right|^{-1}$ or, alternatively, as ${\bar v}^2 < (k+{\tilde c})^{-2}$.

The r.m.s. velocity constraints for the models $M_1^+$, $M_2^+$ and $M_3^-$ determine the range of values of the curvature parameter, $k$, for which linear scaling solutions are allowed, for any given $\beta$ and $N$. Figs. \ref{cc} and \ref{ce} illustrate this in the case of contracting ($M_1^+$ and $M_3^-$) and expanding  ($M_2^+$) models, respectively. The first thing to notice is that, in contracting models, linear scaling regimes are strictly forbidden if ${\tilde c} = 0$ (${\bar v}^2<0$ is not allowed). On the other hand, if ${\tilde c} \neq 0$ the network may  attain linear scaling regimes with velocities within the physically significant range, $\frac{1}{2}<{\bar v}^2<1$. The left and right panels of Fig. \ref{cc} show the allowed range of the curvature parameter, as a function of the expansion exponent, for which the linear scaling solutions are permitted, in the case of the models $M_3^-$ and  $M_1^+$, respectively (for ${\tilde c}=0.5$). In both cases, scaling solutions are allowed for every value of $\beta$ but the allowed range of $k$ is strongly restricted. Also, in both models for ${\tilde c}=0.5$ the causality constraint, given by Eq. (\ref{causality}), does not introduce further restrictions on the plane $(k^2/N,\beta/(1-\beta))$. However, for larger values of ${\tilde c}$, and consequently of $\xi$, the region for which causality is violated widdens. In fact, if the value ${\tilde c}$ is sufficiently large, all linear scaling solutions may be forbidden. In the case of model $M_2^+$, the network is able to reach a linear scaling regime for $ \beta > N^{-1}$ (or equivalently ${\bar v<v_ {\rm max}}$), even if ${\tilde c}=0$, as the left panel of Fig. \ref{ce} illustrates. For larger values of $\beta$, the allowed range of the curvature parameter is only limited by causality. On the other hand, if ${\tilde c} \neq 0$ the linear scaling regime can be attained for any value  of $\beta$, as shown in the right panel of Fig. \ref{ce}.

\subsection{Inflation and superinflation}

In the case of accelerated expansion, as in the models $M_3^+$ (which is inflationary, with $\ddot{a}>0$) and $M_1^-$ (which exhibits superinflation with $\dot{H}>0$), a stretching regime could persist (note that these models would require that ${\tilde c}<0$ in order for a linear scaling solution to be possible). However, in that case, the expansion is fast enough to hamper the brane velocities and make them arbitrarily small. For both models one has
\be
L \propto a \,, \qquad v \propto (Ha)^{-1} \propto a^{-1-1/\beta} \to 0\,.
\ee

In a $(N+1)$-dimensional FRW universe, in which the branes are the dominant component of the energy density, the Einstein equations imply
\bq
\frac{\ddot a}{a} &=& -\frac{8 \pi G_{N+1}}{N(N-1)} \left( (N-2) {\bar \rho} + N {\bar p} \right)\,,\\
\left(\frac{\dot a}{a}\right)^2 &=& \frac{16 \pi G_{N+1}}{N(N-1)} {\bar \rho}\,.
\eq
where $G_{N+1}$ is the $N+1$ dimensional Newton constant. From Eqs. (\ref{em-frw}-\ref{eos}) one obtains
\be
\beta =\frac{2}{N(1+w_b)}= \frac{2}{1+N {\bar v}^2}\,.
\ee
In order to accelerate the universe one needs $\beta >1$ or equivalently $w_b <w_c= (2-N)/N$ and consequently, in a domain wall dominated universe, the r.m.s. velocity of the walls,
\be
{\bar v}^2 < \frac{1}{N}\,,
\ee 
has to be small enough for the universe to be accelerating.

\subsection{Ultra-relativistic collapsing solution}

Consider the case of model $M^-_2$ which represents a collapsing universe with $k<0$ and $0<\beta <1$. A linear scaling solution would only be possible if $k<-{\tilde c}$. However, such solution would necessarily be transient because, if locally the curvature scale of the domain walls becomes smaller than $|H|^{-1}$ then they will tend to become frozen in comoving coordinates while traveling at the speed of light (note that the comoving Hubble radius, $|H|^{-1}$, decreases with time in this model).

Though, in general, in the context of VOS models, the correlation length, $L$, may be identified with the typical physical distance travelled by a brane segment before encountering another segment of the same size, this identification breaks down in the ultrarelativistic limit \cite{Avelino:2002hx,Avelino:2002xy}. In this limit, instead, $L$ is a measure of the energy of the branes and, therefore, the Lorentz factor $\gamma=(1-{\bar v}^2)^{-1/2}$ should be included in the definition of the physical length, $L_{ph}\sim \gamma L$. The fraction of the energy lost by the network due to interface collapse in a timescale $dt$ may be estimated as
\be
-\frac{d {\bar \rho}}{{\bar \rho}}=\frac{dL}{L}\sim \frac{v}{L_{ph}}dt \sim \frac{v}{\gamma L}dt\,.
\ee
Hence, in the $v\rightarrow 1$ limit
\be
{\tilde c} \propto \gamma^{-1} \propto a^{N}\rightarrow 0\,.
\ee

As a consequence, we find that
\be
{\dot L}=\left(1+N\right)HL\,,
\label{vos2u}
\ee
which indicates that the brane network will behave effectivelly as a radiation component. In this limit the domain walls are ultra-relativistic with $\gamma \propto a^{-N}$ but the characteristic length of the network becomes frozen in comoving coordinates so that $L_{ph} \propto a$. The combination of these two effects gives the solution to Eq. (\ref{vos2u}), $L \propto a^{1+N}$.

\section{Friction dominated regime}

If a domain wall moves through a radiation fluid, the interaction with the ultrarelativistic particles results in a frictional force which may be written as  \cite{1994csot.book.....V}
\be
\mathbf{F_f}=-\frac{\sigma}{\ell_{\rm f}} \gamma \mathbf{v}
\ee
where $\ell_{\rm f} \sim \sigma/{\rho_{rad}}\propto a^{N+1}$ is the friction lengthscale and $\rho_{rad}\propto a^{-\left(N+1\right)}$ is the energy density of the relativistic particles.

\subsection{Expanding universe}

As in the case of cosmic strings and domain wall networks in a $(3+1)$-dimensional universe, we expect $(N-1)$-brane networks in a $(N+1)$-dimensional universe to admit two different scaling regimes during the initial part of its evolution, while the friction lengthscale is significantly smaller than the Hubble radius ($\ell_{\rm f}\ll H^{-1}$). In this limit $|{\dot {\bar v}}| \ll {\bar v}/\ell_{\rm f}$ so that ${\bar v}=k \ell_{\rm f}/L$.

If the initial density is sufficiently low ($HL \gg {\bar v}^2 L/\ell_{\rm f}$ and $HL \gg {\tilde c}{\bar v}$) the branes will be conformally stretched by expansion. During this stretching regime the scaling laws are
\be
L \propto a \,, \qquad v \propto \frac{\ell_{\rm f}}{a} \propto a^N \label{friction1}\,.
\ee
This is a transient regime since the velocity increases rather quickly, due to the effect of the domain wall curvature.

As the friction lengthscale approaches the characteristic length of the network, the Kibble regime emerges. Given the fact that, during this regime, the characteristic velocity is higher than it has hitherto been, a considerable amount of energy will be lost due to brane self-interaction ($HL \sim {\tilde c}{\bar v}$). Therefore, even though friction still dominates the dynamics, the scaling laws are different from those of the stretching regime
\bq
L \propto {\sqrt{\frac{\ell_{\rm f}}{|H|}}} \,, \qquad {\bar v} \propto {\sqrt{\ell_{\rm f} |H|}} \label{friction2}\,.
\eq
If the initial density of the network is high enough, the Kibble regime occurs right after the formation of the domain walls and the network will not experience the stretching regime. As the universe expands, the friction lengthscale grows faster than $H^{-1}$ (if $N+1 > 1/\beta$) and will, eventually, overcome the characteristic length of the network, thus bringing the Kibble regime to an end.

\subsection{Collapsing universe}

The scaling solutions given by Eqs.  (\ref{friction1}) and (\ref{friction2}) also account for the dynamics of a domain wall network in a collapsing universe during a friction dominated era. In this case, the domain wall network ends in a friction domination era, coming to a standstill in comoving coordinates and then being conformally contracted (with $L \propto a$). Hence, in this regime the average energy density of the network is given by ${\bar \rho} \propto a^{-1}$. As the background temperature and density approach those of the original wall-forming phase transition, the branes effectively dissolve back into the high density radiation background.

\section{Conclusions}

In this paper we have studied the evolution of domain wall networks, in flat expanding or collapsing homogeneous and isotropic backgrounds with an arbitrary number of spatial dimensions, using a VOS model. We have obtained the scaling laws in frictionless and friction dominated regimes and determined the constraints which have to be satisfied by linear scaling domain wall networks.

The present work is a significant improvement over previous analytical studies of domain wall network evolution, unifying in  a common framework the dynamics of domain wall networks in expanding/collapsing and frictionless/friction dominated regimes. The generalization of the VOS model to $(N+1)$-dimensional FRW universes also provides an important tool to describe the evolution of domain wall networks in more than $3$ spatial dimensions which, up to now was restricted to very special configurations.


\bibliography{branas}

\end{document}